\documentclass[preprint,authoryear,12pt]{elsarticle}
\usepackage{amssymb}
\usepackage{natbib}
\journal{arXiv}

\begin{document}
\begin{frontmatter}
\title{The Inverse Weibull Survival Distribution\\and its Proper Application}
\author{Pasquale Erto}
\address{University of Naples Federico II\\
P.le V. Tecchio, 80\\80125 Naples, Italy}
\begin{abstract}
The peculiar properties of the Inverse Weibull (IW) distribution are shown. 
It is proven that the IW distribution is one of the few models 
having upside-down bathtub (UBT) shaped hazard function. Three real and typical 
degenerative mechanisms, which lead exactly to the IW random variable, are 
formulated. So a new approach to proper application of this 
relatively unknown survival model is supported. However, we consider also 
the case in which any knowledge about generative 
mechanism is unavailable. In this hypothesis, we study a procedure based on 
the Anderson-Darling statistic and log-likelihood function to discriminate 
between the IW model and others alternative UBT distributions. The invariant properties of the 
proposed discriminating criteria have been proven. Based on Monte Carlo simulations, the 
probability of the correct selection has been computed. A real applicative example closes the 
paper.
\end{abstract}
\end{frontmatter}
\section{Introduction}
\label{subsec:introduction} 
Nowadays, the diffused innovation policies require frequent survival 
estimates based on necessarily small samples. That may happen when the 
reliability of technological products -- continuously improved -- must be 
monitored; or when the efficacy of always-new chemotherapy must be promptly 
checked.

In helping statisticians to choose a suitable survival model, careful 
consideration of the generative mechanisms of the involved random variable 
(rv) plays an important (often neglected) role. Such consideration can supplement or even 
prevail over usual model selection procedures, when the observations are
extremely few and, consequently, the information about the effective shape 
of the "parent" distribution (i.e. the population distribution) is very scarce. 

In this context, the paper provides the mathematical models of three typical 
generative mechanisms of the Inverse Weibull (IW) rv. So, the paper helps 
exploiting the IW model to give correct answers for some specific survival 
problems, found in Biometry and Reliability, for which it appears the 
natural interpretative stochastic model.

Doubtless, the IW rv is not widely known and so scarcely identified. The IW 
model is referred to by many different names like ``Frechet-type'' (Johnson 
et al. 1995),  ``Complementary Weibull'' (Drapella 1993), ``Reciprocal Weibull'' (Lu and Meeker 1993; Mudholkar and Kollia 1994), and ``Inverse Weibull'' (Erto 1982; Erto 1989; Johnson et al. 1994; Murthy et al. 2004). An early study of the IW model is reported in the unprocurable paper (Erto 1989). However, it seems to be no comprehensive reference in the literature that studies the IW as survival model. This paper tries to do that specifically exploring its peculiar probabilistic and statistical characteristics. The peculiar heavy right tail of probability density as well as the upside-down bathtub (UBT) shaped hazard function of the IW model has been really found in several applications (Nelson 1990; Rausand and Reinertsen 1996; Gupta et al. 1997; Gupta et al. 1999; Jiang et al. 2003). Also the Inverse Gamma, Inverse Gaussian, Log-Normal, Log-Logistic, and the Birnbaum-Saunders models show similarly shaped hazard rates (Glen 2011; Klein and Moeschberger 2003; Lai and Xie 2006). However, a model incorrectly fitted to IW data may lead to very wrong critical prognoses, even despite its good fitting to the empirical distribution. In fact, especially when few 
observations are available, the empirical distribution contains scarce 
information about the shape of the far-right tail, which is the main and 
unusual feature of the IW distribution. So, the knowledge of primary 
generative mechanisms leading to the IW rv can help one not to miss 
its proper application in some real life peculiar circumstances, analytically shown in the following.

Obviously, the inverse of the IW data follows a Weibull distribution. So the 
parameter estimates of the IW distribution can be easily obtained by 
applying to its reciprocal data the same standard procedures implemented in 
packages for the Weibull model (see Murthy et al. 2004).

\section{Applicative statistical properties}
\label{subsec:statistical}
The probability density function (pdf) of the IW rv $T,$ with scale 
parameter $a$ and shape parameter $b,$ is:
\begin{equation}
\label{eq1}
f(t)=ab(at)^{-(b+1)}\exp \{-(at)^{-b}\},\quad \quad \quad t\ge 0,\quad \quad 
a,\;b>0.
\end{equation}
It is skewed and unimodal for $t_{m} ={\{b \mathord{\left/ {\vphantom {b 
{(b+1)}}} \right. \kern-\nulldelimiterspace} {(b+1)}\}^{1 \mathord{\left/ 
{\vphantom {1 b}} \right. \kern-\nulldelimiterspace} b}} \mathord{\left/ 
{\vphantom {{\{b \mathord{\left/ {\vphantom {b {(b+1)}}} \right. 
\kern-\nulldelimiterspace} {(b+1)}\}^{1 \mathord{\left/ {\vphantom {1 b}} 
\right. \kern-\nulldelimiterspace} b}} a}} \right. 
\kern-\nulldelimiterspace} a$. The $k$th moment of the IW rv is 
$\mbox{E}\{T^{k}\}=1 \mathord{\left/ {\vphantom {1 {a^{k}}}} \right. 
\kern-\nulldelimiterspace} {a^{k}}\Gamma (1-k \mathord{\left/ {\vphantom {k 
b}} \right. \kern-\nulldelimiterspace} b)$ and it exists if $b>k.$ Then the 
mean $\mbox{E}\{T\}=(1 \mathord{\left/ {\vphantom {1 a}} \right. 
\kern-\nulldelimiterspace} a)\Gamma (1-1 \mathord{\left/ {\vphantom {1 b}} 
\right. \kern-\nulldelimiterspace} b)$ and the variance $\mbox{Var}\{T\}=(1 
\mathord{\left/ {\vphantom {1 {a^{2}}}} \right. \kern-\nulldelimiterspace} 
{a^{2}})\{\Gamma (1-2 \mathord{\left/ {\vphantom {2 b}} \right. 
\kern-\nulldelimiterspace} b)-\Gamma^{2}(1-1 \mathord{\left/ {\vphantom {1 
b}} \right. \kern-\nulldelimiterspace} b)\}$ follows.

The most distinctive applicative feature of the IW model is its heavy right tail. That 
is highlighted by the \textit{Property n. 1}: ``The pdf of the IW model is infinitesimal of lower 
order than the negative exponential as $t$ goes to infinity.'' In fact, the ratio of the IW pdf (\ref{eq1}) (setting $a=b=1$, for 
simplicity) to the negative Exponential function goes to infinity as $t$ goes to infinity.

The cumulative distribution function (Cdf) $F(t)$, the survival function 
(Sf) $R(t)$ and the hazard rate (hr) $h(t)$ are easily derived from (\ref{eq1}):
\begin{equation}
\label{eq2}
F(t)=1-R(t)=P(T\le t)=\int_0^t {f(x)dx=} \, \exp \{-(at)^{-b}\}
\end{equation}
\begin{equation}
\label{eq3}
h(t)=\frac{f(t)}{R(t)}=\frac{ab(at)^{-(b+1)}}{\exp 
\{(at)^{-b}\}-1},\quad \quad \quad t>0.
\end{equation}
The hr is infinitesimal as $t$ goes to infinity. It is unimodal and belongs 
to the UBT class (see Glaser 1980) with only one change point: \textit{Property n. 2}: ``The hr of 
the IW model has a unique global maximum between the mode $t_{m} $ and 
the value $t_{n} ={b^{1 \mathord{\left/ {\vphantom {1 b}} \right. 
\kern-\nulldelimiterspace} b}} \mathord{\left/ {\vphantom {{b^{1 
\mathord{\left/ {\vphantom {1 b}} \right. \kern-\nulldelimiterspace} b}} a}} 
\right. \kern-\nulldelimiterspace} a$.''  The condition of maximum for the IW hr does not 
lead to a closed-form solution. However, taking the derivative of the 
logarithm of the IW hr (and appropriately arranging the terms) the necessary 
condition for the maximum of the hr implies that:
\begin{equation}
\label{eq4}
\frac{\exp \{-(at)^{-b}\}}{t}=\frac{1}{t}-\frac{ab(at)^{-(b+1)}}{b+1},\quad \quad \quad t>0.
\end{equation}
The auxiliary functions $U(t)$ and $V(t),$ corresponding to the first and 
second members of this equation, have a unique intersection point. In the 
first quadrant these two functions are both increasing up to their maximum 
point, whose abscissa is for both functions equal to $t_{n} ={b^{1 
\mathord{\left/ {\vphantom {1 b}} \right. \kern-\nulldelimiterspace} b}} 
\mathord{\left/ {\vphantom {{b^{1 \mathord{\left/ {\vphantom {1 b}} \right. 
\kern-\nulldelimiterspace} b}} a}} \right. \kern-\nulldelimiterspace} a$ and 
then they are both decreasing and infinitesimal to the same order as $t$ 
goes to infinity. Moreover, it is possible to verify that $U(t)$ is null as 
$t$ goes to 0, while $V(t)$ is null for the IW mode $t=t_{m} $. Because of the 
following inequalities:
\begin{equation}
\label{eq5}
t_{m} <t_{n} ;\quad \quad U(t_{m} )>V(t_{m} )=0;\quad \quad U(t_{n} 
)<V(t_{n} )
\end{equation}
we derive that the intersection point of the two auxiliary functions, that is 
the maximum point of the hr, falls between the mode $t_{m} ={\{b 
\mathord{\left/ {\vphantom {b {(b+1)}}} \right. \kern-\nulldelimiterspace} 
{(b+1)}\}^{1 \mathord{\left/ {\vphantom {1 b}} \right. 
\kern-\nulldelimiterspace} b}} \mathord{\left/ {\vphantom {{\{b 
\mathord{\left/ {\vphantom {b {(b+1)}}} \right. \kern-\nulldelimiterspace} 
{(b+1)}\}^{1 \mathord{\left/ {\vphantom {1 b}} \right. 
\kern-\nulldelimiterspace} b}} a}} \right. \kern-\nulldelimiterspace} a$ and 
$t_{n} ={b^{1 \mathord{\left/ {\vphantom {1 b}} \right. 
\kern-\nulldelimiterspace} b}} \mathord{\left/ {\vphantom {{b^{1 
\mathord{\left/ {\vphantom {1 b}} \right. \kern-\nulldelimiterspace} b}} a}} 
\right. \kern-\nulldelimiterspace} a$.

The mean residual life ($\mbox{MRL}_{R} $, also called the life expectancy 
of the $R$ fraction of items lived longer than $t_{R} )$ is:
\begin{equation}
\label{eq6}
\begin{array}{c}
 m(t_{R} )=\frac{1}{R(t_{R} )}\int_{t_{R} }^{+\infty } {x\;f(x)dx} -t_{R} = 
\\ 
 =\frac{1 \mathord{\left/ {\vphantom {1 a}} \right. 
\kern-\nulldelimiterspace} a\;\Gamma (1-1 \mathord{\left/ {\vphantom {1 b}} 
\right. \kern-\nulldelimiterspace} b,\;a^{-b}\;t_{R}^{-b})}{1-\exp 
\{-(a\;t_{R} )^{-b}\}}-t_{R} ,\quad \quad b>1 \\ 
 \end{array}
\end{equation}
being $\Gamma (1-1 \mathord{\left/ {\vphantom {1 b}} \right. 
\kern-\nulldelimiterspace} b,\;a^{-b}t^{-b})$ the lower incomplete gamma 
function.

The following \textit{Property n. 3} stands: ``The $\mbox{MRL}_{R} $ function of the IW model is 
bathtub-shaped.'' This property can be deduced from the general results 
given in Gupta and Akman (1995) and is in agreement with the properties of 
the hr. So, the IW model belongs to the class of distribution for which the 
reciprocity of the shape of the hr and $\mbox{MRL}_{R} $ functions holds. 
Specifically, the $\mbox{MRL}_{R} $ decreases from the initial value 
$\mbox{E}(T)$ (as $t$ goes to 0) to its minimum at the change point $t_{0} $ and 
then increases infinitely as $t$ goes to infinity. Being ${d{\kern 1pt}m(t)} 
\mathord{\left/ {\vphantom {{d{\kern 1pt}m(t)} {dt}}} \right. 
\kern-\nulldelimiterspace} {dt}=m(t)h(t)-1$ (e.g., see Lai and Xie, 2006, 
chap. 4), the change point $t_{0} $ must solve the equation $m(t){\kern 
1pt}h(t)=1$ necessarily.

In practice, this peculiar $\mbox{MRL}_{R} $ shape can be found, for 
example, in some biometry problems when the longer the patient's survival 
time from his tumor ablation the better his prognosis.

\section{Real life generative mechanisms}
\label{subsec:generative}
If $T_{1} ,T_{2} ,\mathellipsis ,T_{n} $ are i.i.d. random variables, the 
limit distribution for their maximum is the IW distribution (\ref{eq2}) (Johnson 
\textit{et al.} 1995). Therefore, for instance, when a disease or failure is related to 
the maximum value of a critical non-negative variable, this generative 
mechanism can be considered.

This generative mechanism differs from the following three 
new ones, since for these the time variable does play an explicit role in 
their modeling.

\subsection{``Deterioration'' mechanism}
\label{subsubsec:mylabel1}
Let $Y(t)$ be a system deterioration index that, as such, is a strictly increasing function of the run time $t$. At every intercept with the vertical line passing through $t$, suppose that the uncertainty about $Y(t)$ can be reasonably fitted by a Weibull pdf, with shape parameter constant and scale parameter $u$, function of $t$, modeled by a generic power law:
\begin{equation}
\label{eq8}
u(t)=k\;t^{h},\quad \quad h,\;k>0.
\end{equation}
If a threshold (maximum, positive) value allowed for $Y(t)$ exists, the system 
has the IW Sf. In fact, consider a Weibull random variable $Y(t)$ with 
pdf:
\begin{equation}
\label{eq9}
\begin{array}{r}
 g(y)=v \mathord{\left/ {\vphantom {v {u(t)}}} \right. 
\kern-\nulldelimiterspace} {u(t)}\{y \mathord{\left/ {\vphantom {y {u(t)}}} 
\right. \kern-\nulldelimiterspace} {u(t)}\}^{v-1}\exp [-\{y \mathord{\left/ 
{\vphantom {y {u(t)}}} \right. \kern-\nulldelimiterspace} {u(t)}\}^{v}], \\ 
 y\ge 0,\quad v,\;u>0 \\ 
 \end{array}
\end{equation}
where $v$, the shape parameter, is constant, and $u(t)$, the scale 
parameter, is the drift function (\ref{eq8}). If $D$ is the threshold (maximum, 
positive) value for $Y(t)$, then:
\begin{equation}
\label{eq10}
R(t)=P\{Y(t)<D\}=\int_0^D {g(y)dy=1-\exp [-\{D \mathord{\left/ {\vphantom {D 
{u(t)}}} \right. \kern-\nulldelimiterspace} {u(t)}\}^{v}]} .
\end{equation}
Substituting $u(t)=k\,t^{h}$ back into the previous relationship, we obtain:
\begin{equation}
\label{eq11}
R(t)=1-\exp [-\{(k \mathord{\left/ {\vphantom {k D}} \right. 
\kern-\nulldelimiterspace} D)^{1 \mathord{\left/ {\vphantom {1 h}} \right. 
\kern-\nulldelimiterspace} h}t\}^{-vh}].
\end{equation}
On putting $a=(k \mathord{\left/ {\vphantom {k D}} \right. 
\kern-\nulldelimiterspace} D)^{1 \mathord{\left/ {\vphantom {1 h}} \right. 
\kern-\nulldelimiterspace} h}$ and $b=v\,h,$ the IW Sf follows.

This mechanism is found in many technological corrosion phenomena that give 
rise to failures only when they reach a threshold deepness $D.$ The 
mechanism is found also in many biologic degenerative phenomena (i.e., 
gradual deterioration of organs and cells) where the loss of function 
appears when the deterioration deep $Y(t)$ reaches a fixed threshold value. 
Besides, this mechanism is found when tumors spread potential metastases 
with a dissemination probability proportional to their size $Y(t)$. Hence, 
a tumor size greater than a given threshold value $D$ causes a rate of 
occurrence of metastases which is really first increasing and then 
decreasing (see Le Cam and Neyman 1982, p. 253) like the IW one (\ref{eq3}).

\subsection{``Stress-Strength'' mechanism}
\label{subsubsec:mylabel2}
If the stress $S$ (in the broad sense) is a rv with distribution that can be 
reasonably fitted by a Weibull model and the strength $Z,$ that opposes $S,$ 
is a decreasing function of time $t$ that can be modeled by a generic power 
law:
\begin{equation}
\label{eq12}
Z(t)=k\,t^{-h},\quad \quad h,\;k>0
\end{equation}
the resulting Sf is the IW one. In fact, if the stress $S$ is a Weibull random variable:
\begin{equation}
\label{eq13}
g(s)=v \mathord{\left/ {\vphantom {v u}} \right. \kern-\nulldelimiterspace} 
u(s \mathord{\left/ {\vphantom {s u}} \right. \kern-\nulldelimiterspace} 
u)^{v-1}\exp \{-(s \mathord{\left/ {\vphantom {s u}} \right. 
\kern-\nulldelimiterspace} u)^{v}\}\,,\quad \quad s\ge 0,\quad u,\;v>0
\end{equation}
and the strength $Z,$ that opposes $S,$ follows the decreasing function of 
time (\ref{eq12}):
\begin{equation}
\label{eq14}
\begin{array}{c}
 R(t)=P\{S<Z(t)\}=\int_0^{Z(t)} {g(s)ds} = \\ 
 =1-\exp [-\{{Z(t)} \mathord{\left/ {\vphantom {{Z(t)} u}} \right. 
\kern-\nulldelimiterspace} u\}^{v}]. \\ 
 \end{array}
\end{equation}
Substituting $Z(t)=k\,t^{-h}$ back into the previous relationship, we 
obtain:
\begin{equation}
\label{eq15}
R(t)=1-\exp [-\{(u \mathord{\left/ {\vphantom {u h}} \right. 
\kern-\nulldelimiterspace} h)^{1 \mathord{\left/ {\vphantom {1 h}} \right. 
\kern-\nulldelimiterspace} h}t\}^{-vh}]
\end{equation}
then, renaming $a=(u \mathord{\left/ {\vphantom {u h}} \right. 
\kern-\nulldelimiterspace} h)^{1 \mathord{\left/ {\vphantom {1 h}} \right. 
\kern-\nulldelimiterspace} h}$ and $b=v\,h,$ the IW Sf follows.

This mechanism is common for many mechanical components (see, for example, 
Bury 1975, p. 593; Shigley 1977, p. 184) as well as it is found in patients 
with a decreasing vital strength following the (\ref{eq12}) (e.g., because they are 
subjected to intensive and prolonged chemotherapy) and subjected to a 
relapse having a random virulence or gravity $S.$ In these cases, an hr 
first quickly increasing and then slowly decreasing, is sometimes 
surprisingly observed (see Carter et al. 1983, p. 79).

\subsection{``Unsuccessful-Defensive-Attempts'' mechanism}
\label{subsubsec:mylabel3}
Suppose that a disease (or failure) is latent and the physiological 
defensive attempts averse to it occur randomly according to a Poisson model. 
If the probability of one successful defensive attempt depends on the 
incubation time $t$ (but not on the number of previously occurred defensive 
actions) according to a generic power law decreasing function:
\begin{equation}
\label{eq16}
P_{S} (t)=k\,t^{-h},\quad \quad h>1,\quad k>0,\quad t\ge k^{\frac{1}{h}}
\end{equation}
the IW Cdf follows. In fact, suppose that the random variable $N_{a} ,$ 
describing the physiological defensive attempts against a latent disease (or 
failure), occurs according to a Poisson law:
\begin{equation}
\label{eq17}
\begin{array}{r}
 P(N_{a} =n_{a} )=\{{(\beta t)^{n_{a} }} \mathord{\left/ {\vphantom {{(\beta 
t)^{n_{a} }} {n_{a} !}}} \right. \kern-\nulldelimiterspace} {n_{a} !}\}\exp 
(-\beta t), \\ 
 n_{a} =0,1,2,\mathellipsis ,\quad \quad \beta >0. \\ 
 \end{array}
\end{equation}
Let $P_{S} $ be the probability of one successful defensive attempt, which 
depends on the incubation time $t$ (but not on the number of previously 
occurred defensive actions) according to the function (\ref{eq16}). Consequently, 
the probability of manifest disease (or failure) is:
\begin{equation}
\label{eq18}
\begin{array}{l}
 F(t)=\exp (-\beta t)\{1+(\beta t)(1-kt^{-h})+ \\ 
 +\frac{(\beta t)^{2}}{2!}(1-kt^{-h})^{2}+\mathellipsis \}=\exp \{-\beta 
kt^{-(h-1)}\}. \\ 
 \end{array}
\end{equation}
Then, on putting $b=h-1$ and $a=(\beta k)^{-1 \mathord{\left/ {\vphantom {1 
b}} \right. \kern-\nulldelimiterspace} b},$ the IW Cdf follows.

This mechanism is found in Biometry when the immune system works randomly 
against antigens, and its effectiveness decreases as the disease expands 
(see Le Cam and Neyman 1982, p. 15). In reliability, this mechanism is found 
when a technological system is randomly (i.e., without any definite plan) 
maintained: the smaller the time from the beginning of the failure process 
(up to the maintenance action) the greater the maintenance efficacy.

\section{The problem of the IW model selection}
\label{subsec:first}
Consider the following 50 pseudo random (ordered) data generated from a ``close-to-standard" parent Cdf 
(\ref{eq2}) with $a=1$ and $b=1.1$ (we cannot put $b=1$ since, in general, the  $k$th moment of the IW pdf exists if  $b>k$) :

0.2776, 0.2931, 0.3384, 0.4321, 0.4739, 0.4771, 0.5331, 0.5424, 0.5482, 
0.5571, 0.6139, 0.6451, 0.6523, 0.6587, 0.7166, 0.7838, 0.8466, 0.8892, 
0.9278, 0.9651, 1.008, 1.051, 1.123, 1.203, 1.213, 1.366, 1.529, 1.795, 
1.947, 2.093, 2.143, 2.189, 2.246, 2.453, 2.526, 2.858, 2.924, 3.381, 3.383, 
3.587, 4.964, 5.101, 5.139, 6.753, 10.11, 11.37, 12.68, 16.88, 17.25, 19.07.

The Anderson-Darling statistic (Anderson and Darling 1954) $A_{n}^{2} 
=0.2927$, with a $p$-value equal to 0.94333, shows the high conformity of this 
sample to the parent Cdf. Incidentally, in this paper, we chose this 
specific goodness-of-fit test since it emphasizes the tails of the presumed 
parent distribution. 
However, in the above case, also tests that give less weight to the tails 
lead to similar results.

Suppose that we want to identify a generic Cdf model being very well fitted to both the data and the parent Cdf, but we don't have any strong information about the latter. We decide to adopt a 
``less informative model'' which is coherent with our poor information. We 
chose a polynomial cumulative hr (Hr) model of order 3, since it is the 
minimum able to fit a non-monotone model too. In our (simulated) condition, we 
can define an excellent ``a priori'' model by fitting the polynomial to 50 
points (vertically equally spaced) of the known parent Cdf. The resulting 
model is:
\begin{equation}
\label{eq19}
H(t)=\int_0^t {h(z)dz} =0.5305\,t-0.03597\,t^{2}+0.0008995\,t^{3},\quad 
\quad h(t)>0
\end{equation}
which has a coefficient of determination $\rho_{d}^{2} =0.9908$. Moreover, 
being the Anderson-Darling statistic $A_{n}^{2} =1.152$, with a $p$-value equal 
to 0.2856, this ``a priori'' model appears very well fitted to data too.
Incidentally, the maximum likelihood (ML) estimates of its three parameters give 
the following polynomial Hr model very close to the former (\ref{eq19}):
\begin{equation}
\label{eq20}
H(t)=0.5427\,t-0.04931\,t^{2}+0.001728\,t^{3},\quad \quad h(t)>0
\end{equation}
which has a coefficient of determination $\rho_{d}^{2} =0.9758$.

Suppose now that the analysis of the generative mechanism suggests us to fit 
the IW model to the 50 data. The ML estimates of its 
parameters are $\hat{{a}}=1.027$ and $\hat{{b}}=1.105$. The coefficient of 
determination of the Hr function estimated from this IW model is $\rho 
_{d}^{2} =0.9648$. The Anderson-Darling statistic is $A_{n}^{2} =0.2740$ 
with a $p$-value equal to 0.9530. 

Although the previous analysis has shown that the two Cdf models fit the data 
very well, some important characteristics could be different. To highlight that, we compare some 
critical estimates obtained from the ``a priori and less informative'' model 
(\ref{eq19}) with those obtained using the last ``fitted and informative'' IW model. 
From these two models we obtain the $\mbox{MRL}_{R} $ estimates reported in \tablename~\ref{tab1}, where the true values are those of the parent population.

\begin{table}[htbp]
\begin{center}
\caption{$\mbox{MRL}_{R} $ estimates for the polinomial and IW fitted models}
\begin{tabular}{|l|l|l|l|}
\hline
\textit{}& 
\textit{polinomial}& 
\textit{IW}& 
\textit{true} \\
\hline
$\mbox{MRL}_{0.50} $& 
{4.268}& 
{17.77}& 
{18.85} \\
\hline
$\mbox{MRL}_{0.25} $& 
{5.833}& 
{33.47}& 
{35.31} \\
\hline
$\mbox{MRL}_{0.10} $& 
{8.958}& 
{77.13}& 
{81.15} \\
\hline
\end{tabular}
\label{tab1}
\end{center}
\end{table}

These results show that the empirical fitting of a model to the IW data can 
lead to wrong model and its effect can be quite severe. So the necessity of 
a suitable strategy to choose the best model among all that (reasonably well) fit the data arises. 
\section{Comparing the Inverse Weibull with other commonly-known distributions}
\label{subsec:comparing}
The above illustrative example is worth only to identify a specific 
goodness-of-fit problem and to promote further studies since, even 
remarkable, its results are obtained without considering other 
heavy-tail-type distributions and they are based on a single draw of 50 
observations.

\begin{figure}[htbp]
\centerline{\includegraphics[width=5.00in,height=3.00in]{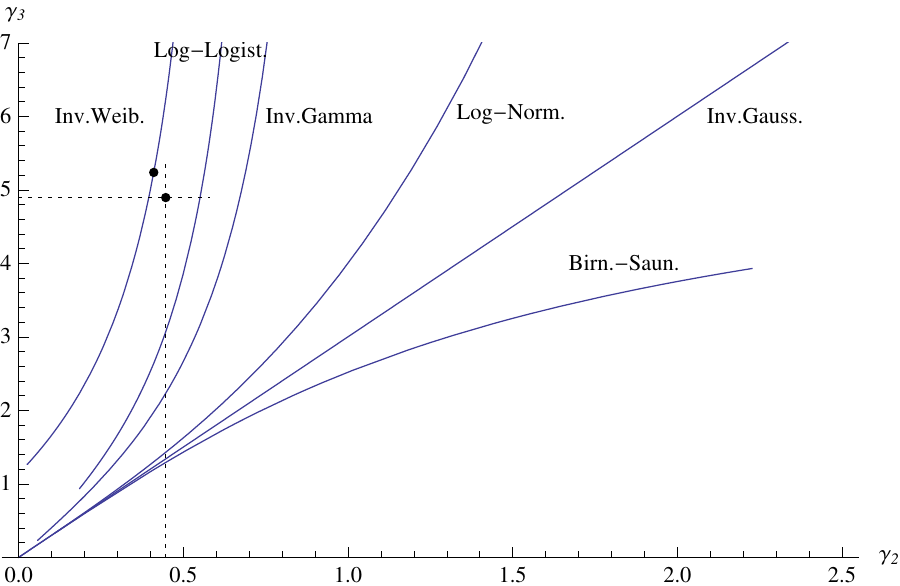}}
\caption{. Coefficient of variation $\gamma_{2} $ against skewness $\gamma_{3} $ for 
various survival models}
\label{fig1}
\end{figure}

To compare the IW model with other potential alternative and 
commonly-known distributions, the chart from (Glen 2011; Vargo et al. 2010) is drawn in 
\figurename~\ref{fig1} including the IW together with the other few models having upside-down 
bathtub (UBT) shaped hazard function.
In this chart, the coefficient of 
variation $\gamma_{2} =\sigma \mathord{\left/ {\vphantom {\sigma \mu }} 
\right. \kern-\nulldelimiterspace} \mu $ is plotted against skewness $\gamma 
_{3} ={\mbox{E}\left\{ {(X-\mu )^{3}} \right\}} \mathord{\left/ {\vphantom 
{{\mbox{E}\left\{ {(X-\mu )^{3}} \right\}} {\sigma^{3}}}} \right. 
\kern-\nulldelimiterspace} {\sigma^{3}}$ for five alternative distribution 
models. Skewness is used to comparatively measure the tendency for one of 
their tails to be heavier than the other. The plot usually includes all 
possible pairs $(\gamma_{2} ,\;\gamma_{3} )$ that a model can attain. The 
set of values that the IW $(\gamma_{2} ,\;\gamma_{3} )$ pairs can assume 
fall at left of those of all the other models, helping to fill a gap on the 
extreme left of the chart. Since it occupies a small part of the chart, the 
IW model confirms the fact that only peculiar data, corresponding to a small 
subset of the allowable moment pairs, can be modeled by it.

Unfortunately, when we have to analyze a sample data, the plot of the sample point 
$(\hat{{\gamma }}_{2} ,\;\hat{{\gamma }}_{3} )$ on such a graph could not show the 
feasible closest models to the data to start the selection. In fact, the sample
skewness is rather an unreliable estimator of the corresponding population parameter 
when the sample size is small (say less than 50). Usually it is 
underestimated, and the bias becomes negligible only for very large sample 
size (say greater than 1000).

Consider the following 50 pseudo random (ordered) data generated from the 
parent Cdf (\ref{eq2}) with $a=1$ and $b=4.1$ (for these values both coefficient of 
variation and skewness of the IW parent distribution exist):

0.7228, 0.7955, 0.8202, 0.8333, 0.8535, 0.8641, 0.8650, 0.9124, 0.9245, 
0.9300, 0.9598, 0.9706, 1.017, 1.017, 1.031, 1.033, 1.047, 1.052, 1.059, 
1.083, 1.102, 1.121, 1.150, 1.152, 1.156, 1.158, 1.175, 1.183, 1.187, 1.203, 
1.204, 1.211, 1.218, 1.226, 1.247, 1.270, 1.305, 1.320, 1.338, 1.347, 1.356, 
1.359, 1.365, 1.389, 1.473, 1.567, 1.637, 1.823, 1.897, 4.637.

The Anderson-Darling statistic $A_{n}^{2} =1.460$, with a $p$-value equal to 
0.1864, shows the conformity of this sample to the parent Cdf. The sample 
point $(\hat{{\gamma }}_{2} ,\;\hat{{\gamma }}_{3} )$ is (0.4464, 4.894) (on 
the cross of the dashed lines in \figurename~\ref{fig1}) and the 
parent distribution point $(\gamma_{2} ,\;\gamma_{3} )$ is (0.4100, 5.236) 
(on the IW curve in \figurename~\ref{fig1}).
Even if the size of the sample is not very high, we consider that the 
plot of the sample point $(\hat{{\gamma }}_{2} ,\;\hat{{\gamma }}_{3} )$ of 
\figurename~\ref{fig1} suggests us to fit the IW and the 
Log-Logistic models to the 50 data, being the Cdf of the latter model:
\begin{equation}
\label{eq21}
F(t)=\, \frac{1}{1+(t \mathord{\left/ {\vphantom {t \sigma }} \right. 
\kern-\nulldelimiterspace} \sigma )^{-\gamma }},\quad \quad \quad t>0,\quad 
\quad \gamma ,\;\sigma >0
\end{equation}
The ML estimates of the IW parameters are $\hat{{a}}=0.9629$ 
and $\hat{{b}}=4.752$, and the Anderson-Darling statistic is $A_{n}^{2} 
=0.5994$ with a $p$-value equal to 0.1250. The ML estimates of 
the Log-Logistic parameters are $\hat{{\sigma }}=1.145$ and $\hat{{\gamma 
}}=7.394$, and the Anderson-Darling statistic is $A_{n}^{2} =0.3587$ with a 
$p$-value equal to 0.3875.
\begin{figure}[htbp]
\centerline{\includegraphics[width=5.00in,height=3.00in]{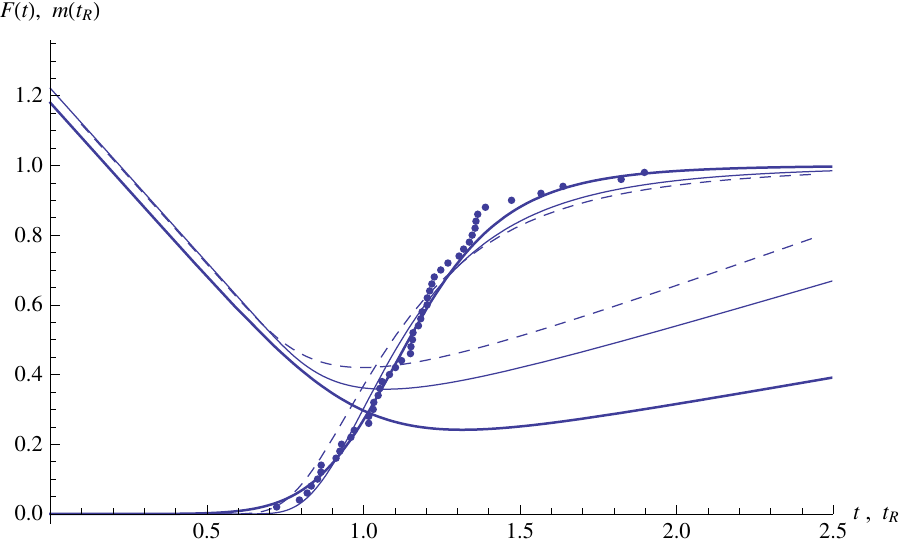}}
\caption{. Cdf and $\mbox{MRL}_{R} $ of the parent (dashed lines) Inverse Weibull 
(thin lines) Log-Logistic (thick lines) models and the sample Cdf points.}
\label{fig2}
\end{figure}
Despite the fact the two models are both well fitted to the data (and appear very close in \figurename~\ref{fig1}) the effect on critical prognoses, of the mis-specification, is remarkable. For example, from the IW model we estimate $\mbox{MRL}_{0.1} =0.4729$ and from the LL model we estimate $\mbox{MRL}_{0.1} =0.2775$ (being 0.5754 the true value.) In fact, the \figurename~\ref{fig2} shows that although the Cdfs of the two 
models are quite close to each other, their $\mbox{MRL}_{R} $ functions are 
rather different. So, we must try to understand how we can select the correct model.

Glen and Leemis (1997) showed that comparisons among many survival distributions can be successfully made by using a goodness-of-fit statistic at its ML value. So, a first strategy could select that distribution with 
the smallest Anderson-Darling statistic at its ML value. The strategy could be refined by considering the difference of the maximized log-likelihoods (\textit{MLLs}) and choosing the distribution with the largest value.
However, comparing the two above alternative models by the Anderson-Darling statistic would lead to incorrect selection, since 
the fitted Log-Logistic model has the smaller statistic $A_{n}^{2} $. Vice 
versa, comparing the two models by means of their \textit{MLLs} would lead to correct 
selection, since the \textit{MLL} of the fitted Log-Logistic 
model is equal to $-9.403$ and the \textit{MLL} of the fitted IW is equal to $-8.134$ 
(although the difference is only 1.269).

The obvious question is: how reliable are both the criteria?

\section{Some properties of the P-AD and P-MLL discriminant indices}
\label{subsec:mylabel1}
We decide to estimate the probabilities of correct selection in terms of the 
fraction of times (\textit{P-AD}) that the fitted IW model has the smaller statistic 
$A_{n}^{2} $ and the fraction of times (\textit{P-MLL}) that the fitted IW model has the 
larger \textit{MLL}. We found that for the IW and Log-Logistic distributions both indices \textit{P-AD }and \textit{P-MLL} are pivotal 
quantities that is independent of the hypothetical distribution parameters 
(intended as ``arbitrary but determined'' values).

\subsection{Pivotal property of the P-AD index}
\label{subsubsec:pivotal}
The Anderson-Darling statistic (Anderson and Darling 1954) used to estimate 
the \textit{P-AD} is:
\begin{equation}
\label{eq22}
A_{n}^{2} =n\int_{-\infty }^\infty {\frac{[F_{n} 
(t)-\hat{{F}}(t)]^{2}}{\sqrt {\hat{{F}}(t)\;[1-\hat{{F}}(t)]} 
}d\hat{{F}}(t)} 
\end{equation}
where $\hat{{F}}(t)$ is the hypothetical Cdf at its ML 
values, based on a sample of size $n$, and $F_{n} (t)$ is the empirical 
Cdf defined as $i \mathord{\left/ {\vphantom {i n}} \right. 
\kern-\nulldelimiterspace} n$ if $i$ of the $t_{1} ,\mathellipsis ,t_{n} $ 
sample data are $\le t$. As already said before, being the parameters of the 
hypothetical Cdf estimated from the data, the distribution of the 
statistic (\ref{eq22}) is evaluated via Monte Carlo simulation based every time upon 
1000 pseudo-random samples.

We begin showing the \textit{Property A}: ``For the IW model the distribution of the maximum 
likelihood estimator $\hat{{F}}(t)$ depends only upon $F(t)$ (\ref{eq2}) and $n$''. 

From (\ref{eq2}) we obtain $t=a^{-1}\;\{-\ln [F(t)]\}^{-1 \mathord{\left/ 
{\vphantom {1 b}} \right. \kern-\nulldelimiterspace} b}$ that inserted in 
$\hat{{F}}(t)$ gives:
\begin{equation}
\label{eq23}
\hat{{F}}(t)=\left[ {-\left( {{\hat{{a}}} \mathord{\left/ {\vphantom 
{{\hat{{a}}} a}} \right. \kern-\nulldelimiterspace} a} 
\right)^{-\hat{{b}}}\left\{ {-\ln \left( {F(t)} \right)^{{\hat{{b}}} 
\mathord{\left/ {\vphantom {{\hat{{b}}} b}} \right. 
\kern-\nulldelimiterspace} b}} \right\}} \right]
\end{equation}
where $\hat{{a}}$ and $\hat{{b}}$ are the maximum likelihood estimates, of 
the respective parameters, and both the quantities $({\hat{{a}}} 
\mathord{\left/ {\vphantom {{\hat{{a}}} a}} \right. 
\kern-\nulldelimiterspace} a)^{-\hat{{b}}}$ and ${\hat{{b}}} \mathord{\left/ 
{\vphantom {{\hat{{b}}} b}} \right. \kern-\nulldelimiterspace} b$ are 
pivotal. In fact, by letting $x=\ln (1 \mathord{\left/ {\vphantom {1 t}} 
\right. \kern-\nulldelimiterspace} t)$ we obtain a location-scale rv $x$, 
with location parameter $l=\ln (a)$ and scale parameter $s=1 \mathord{\left/ 
{\vphantom {1 b}} \right. \kern-\nulldelimiterspace} b$. For all the 
location-scale rv's the quantities${(\hat{{l}}-l)} \mathord{\left/ 
{\vphantom {{(\hat{{l}}-l)} {\hat{{s}}}}} \right. \kern-\nulldelimiterspace} 
{\hat{{s}}}$ and ${\hat{{s}}} \mathord{\left/ {\vphantom {{\hat{{s}}} s}} 
\right. \kern-\nulldelimiterspace} s$ are pivotal (Lawless 2003) being 
$\hat{{l}}$ and $\hat{{s}}$ the maximum likelihood estimates of the 
parameters $l$ and $s$ respectively. Since $({\hat{{a}}} \mathord{\left/ 
{\vphantom {{\hat{{a}}} a}} \right. \kern-\nulldelimiterspace} 
a)^{-\hat{{b}}}=\exp \{-{(\hat{{l}}-l)} \mathord{\left/ {\vphantom 
{{(\hat{{l}}-l)} {\hat{{s}}}}} \right. \kern-\nulldelimiterspace} 
{\hat{{s}}}\}$ and ${\hat{{b}}} \mathord{\left/ {\vphantom {{\hat{{b}}} b}} 
\right. \kern-\nulldelimiterspace} b=({\hat{{s}}} \mathord{\left/ {\vphantom 
{{\hat{{s}}} s}} \right. \kern-\nulldelimiterspace} s)^{-1}$, also both 
these are pivotal quantities and from (\ref{eq23}) it follows the \textit{Property A}.

Now we show the \textit{Property B}: ``For the Log-Logistic model the distribution of the 
maximum likelihood estimator $\hat{{F}}(t)$ depends only on $F(t)$ (\ref{eq21}) and 
$n$''.

From (\ref{eq21}) we obtain $t=\sigma \;\{F(t)^{-1}-1\}^{-1 \mathord{\left/ 
{\vphantom {1 \gamma }} \right. \kern-\nulldelimiterspace} \gamma }$ that 
inserted in $\hat{{F}}(t)$ gives:
\begin{equation}
\label{eq24}
\hat{{F}}(t)=\left[ {\{F(t)^{-1}-1\}^{{\hat{{\gamma }}} \mathord{\left/ 
{\vphantom {{\hat{{\gamma }}} \gamma }} \right. \kern-\nulldelimiterspace} 
\gamma }({\hat{{\sigma }}} \mathord{\left/ {\vphantom {{\hat{{\sigma }}} 
\sigma }} \right. \kern-\nulldelimiterspace} \sigma )^{\hat{{\gamma }}}} 
\right]^{-1}
\end{equation}
where $\hat{{\gamma }}$ and $\hat{{\sigma }}$ are the maximum likelihood 
estimates and both the quantities $({\hat{{\sigma }}} \mathord{\left/ 
{\vphantom {{\hat{{\sigma }}} \sigma }} \right. \kern-\nulldelimiterspace} 
\sigma )^{\hat{{\gamma }}}$ and ${\hat{{\gamma }}} \mathord{\left/ 
{\vphantom {{\hat{{\gamma }}} \gamma }} \right. \kern-\nulldelimiterspace} 
\gamma $ are pivotal. In fact, by letting as before $x=\ln (1 
\mathord{\left/ {\vphantom {1 t}} \right. \kern-\nulldelimiterspace} t)$ we 
obtain a location-scale rv $x$, with location parameter $l=\ln (\sigma )$ 
and scale parameter $s=1 \mathord{\left/ {\vphantom {1 \gamma }} \right. 
\kern-\nulldelimiterspace} \gamma $. Since $({\hat{{\sigma }}} 
\mathord{\left/ {\vphantom {{\hat{{\sigma }}} \sigma }} \right. 
\kern-\nulldelimiterspace} \sigma )^{\hat{{\gamma }}}=\exp 
\{-{(\hat{{l}}-l)} \mathord{\left/ {\vphantom {{(\hat{{l}}-l)} {\hat{{s}}}}} 
\right. \kern-\nulldelimiterspace} {\hat{{s}}}\}$ and ${\hat{{\gamma }}} 
\mathord{\left/ {\vphantom {{\hat{{\gamma }}} \gamma }} \right. 
\kern-\nulldelimiterspace} \gamma =({\hat{{s}}} \mathord{\left/ {\vphantom 
{{\hat{{s}}} s}} \right. \kern-\nulldelimiterspace} s)^{-1}$, also these are 
pivotal quantities and from (\ref{eq24}) it follows the \textit{Property B}.

From the properties $A$ and $B$ it follows the \textit{Property C}: ``The comparison between the 
Anderson-Darling statistics calculated respectively for the fitted IW and 
Log-Logistic models is independent of the hypothetical distribution parameters''. 
This implies the pivotal property of the \textit{P-AD} index.

\subsection{Pivotal property of the P-MLL index}
\label{subsubsec:mylabel4}
From the properties $A$ and $B$, it follows that for both IW and Log-Logistic 
models the pdf (and so the log-likelihood) calculated at its maximum 
likelihood values, is independent of distribution parameters. Consequently, 
the same property is valid for the comparison between their maximized 
log-likelihoods.

\subsection{Estimates of the P-AD and P-MLL indices}
\label{subsubsec:about}
For every combination of values $a=(1, 2, 3)$, $b=(1.1, 2.1, 3.1, 4.1, 5.1)$ and $n=(10, 30, 50)$, we generated 1000 pseudo random samples 
from the parent IW distribution and computed \textit{P-AD}, \textit{P-MLL} and the fraction of times 
(\textit{P-AD{\&}MLL}) that the fitted IW model has both the smaller statistic $A_{n}^{2} $ and 
the larger \textit{MLL}.

Thanks to the pivotal property of the \textit{P-AD} and \textit{P-MLL} indices,  
the conducted simulations gave 15 nearly identical results for each $n$. So, we have been able to evaluate a very reliable estimate 
of the probability of correct model selection (\tablename~\ref{tab2}) based on the three examined criteria respectively. 
It is evident that \textit{P-MLL} includes \textit{P-AD} -- in terms of fraction of times of correct selection -- 
and that the selection of the fitted model based upon the larger \textit{MLL }has the 
highest probability of being correct.

\begin{table}[htbp]
\begin{center}
\caption{Probability of correct model selection estimated by averaging 15000 simulated results}
\begin{tabular}{|l|l|l|l|}
\hline
$n$& 
\textit{P-AD}& 
\textit{P-MLL}& 
\textit{P-AD{\&}MLL} \\
\hline
10& 
0.60& 
0.78& 
0.78 \\
\hline
30& 
0.77& 
0.88& 
0.88 \\
\hline
50& 
0.85& 
0.93& 
0.93 \\
\hline
\end{tabular}
\label{tab2}
\end{center}
\end{table}

\section{Times to Breakdown of a Capacitor Insulating Fluid}
This example is representative of the critical real-world situations in which 
only tiny data sets are available. The dataset consists of 15 times to 
breakdown (in minutes) of an insulating fluid between electrodes at a 
constant voltage $V$ (36 kV), provided in Nelson (1982, p. 105):

\begin{center}
0.35, 0.59, 0.96, 0.99, 1.69, 1.97, 2.07, 2.58, 2.71, 2.90, 3.67, 3.99, 
5.35, 13.77, 25.50.
\end{center}
Unfortunately, due to small size of the sample, we cannot rely on the 
sample point $(\hat{{\gamma }}_{2} =1.439,\;\;\hat{{\gamma }}_{3} =2.428)$ 
on the graph of \figurename~\ref{fig1} to start the selection of a 
reasonable model. 

However, analyzing the experiment (aiming to derive the lifetime 
distribution of the insulating fluid) we come to the conclusion that it 
shows an example of the ``Deterioration'' mechanism close to the one 
described in Section 3.1. In fact, the mean of the insulating resistance 
$\Omega $ of the fluid decreases according to a positive (and less than 
one) power function of time. This model belongs to the Arrhenius class of 
cumulative damage relationships, widely found in life tests with constant 
stress (see, e.g., Nelson 1990). Consequently, the mean of the resistive 
leakage current $I\cong V \mathord{\left/ {\vphantom {V \Omega }} \right. 
\kern-\nulldelimiterspace} \Omega $ (i.e., the system deterioration index 
$Y)$ increases with a positive (and greater than one) power of time to 
the dielectric failure, which occurs when a threshold value $D$ (fixed by the 
operating and environmental conditions supposed constant) is exceeded. 
Moreover, the nature of the failure mechanism is stationary and does not 
induce any change in the shape of the $Y$ pdf. Then, a pdf model -- with 
mean increasing as a power function of time and with constant shape -- is 
well rendered by the Weibull model (\ref{eq9}). In fact, being constant the shape 
parameter $v$, its mean $u(t)\;\Gamma (1 \mathord{\left/ {\vphantom {1 v}} 
\right. \kern-\nulldelimiterspace} v+1)$ is effectively a positive (and 
greater than one) power function of the time.

Hence we decide to assume the IW model as our weighted hypothesis. However, 
we consider also the Log-Logistic model because, as shown in 
\figurename~\ref{fig1}, it plays the role of a frontier 
separating the IW model and many other alternative models.

The ML estimates of the IW parameters are $\hat{{a}}=0.688$ 
and $\hat{{b}}=1.03$; the Anderson-Darling statistic is $A_{n}^{2} =0.312$ 
with a $p$-value equal to 0.596; the \textit{MLL} is $MLL=-36.1$. 
The ML estimates of the Log-Logistic parameters are 
$\hat{{\sigma }}=1.68$ and $\hat{{\gamma }}=2.37$ and the Anderson-Darling 
statistic is $A_{n}^{2} =0.201$ with a $p$-value equal to 0.870; the maximized 
log-likelihood is $MLL=-35.8$. The comparison of the two alternative models 
by means of the Anderson-Darling statistic and the \textit{MLLs}
(both at their ML value) would support the Log-Logistic 
model. However, we think that the differences are not enough large (e.g. only 
0.3 unit separates the two \textit{MLLs}) to contradict the 
previous choice based on a careful and detailed technological analysis.

\section{Concluding remarks}
\label{subsec:concluding}
The paper proves that the IW distribution is another of the 
relatively few UBT survival distributions. So, when dealing with UBT 
distributions, it is helpful to have an alternative model that has, moreover, 
a distinctive heavy right tail.

This paper demonstrates how the IW distribution is the natural 
candidate, among all the survival models, to face three unreported classes 
of real and well defined degenerative phenomena. So the practitioners are helped to 
choose this model by profiting from the knowledge of the involved phenomena, 
such as a disease or failure, rather than exclusively on the usual analysis 
of goodness-of-fit.

Some illustrative examples show that the polynomial cumulative hazard 
model and the Log-Logistic one can both fit the Cdf of IW data very well. 
The polynomial model is used as antithetic benchmark because: a) differently 
from the IW model, it is capable of giving a wide range of hr shapes; b) it 
is used in situations where strong assumptions about the parent distribution are 
unavailable. The Log-Logistic model has been considered because: a) it is the closest model which 
shares the upside-down bathtub (UBT) shaped hazard 
function; b) it plays the role of a frontier separating the IW model 
from many other alternative models.

However, all the illustrative examples show that the above models -- even 
though very well fitted to IW data -- may be very misleading because they entail 
highly incorrect assessments concerning, for instance, the mean residual life.

The paper proves that -- when any knowledge about generative 
mechanism is unavailable -- selecting between the IW and the Log-Logistic 
models that one which minimizes the Anderson-Darling statistic or, even 
better, maximizes the likelihood is a very effective procedure.

We found that the correct selection based on the Anderson-Darling 
statistic implies that one based on the maximized log-likelihood, but the 
vice versa is not true.

Finally, we show that for the IW and Log-Logistic models both selection 
criteria are independent of hypothetical distribution parameters, and the 
corresponding probabilities of correct selection are respectively greater 
than 0.85 and 0.93 when the size of the available sample is greater than 50. 
Instead, when the size of the available sample is less than 30 (i.e., in a 
very frequent situation in the technological and biological fields) 
selecting the correct model purely on the basis of the empirical 
distribution remains a highly risky procedure, since the probabilities of 
wrong selection are respectively greater than 0.23 and 0.12.\\

\end{document}